\documentclass[a4paper]{jpconf}

\usepackage[utf8]{inputenc}
\usepackage[english]{babel}
\usepackage{iopams}
\usepackage{xspace}
\usepackage{environ}
\usepackage{mathtools}
\usepackage{physics}

\newcommand{\ie}{{i.e.}\xspace}

\newcommand{\fm}{\ensuremath{\mathrm{fm}}}

\DeclareMathOperator*{\argmin}{arg\,min}

\newcommand{\mbraket}[3]{\langle #1|#2|#3\rangle}

\newcommand{\ii}{\mathrm{i}}

\NewEnviron{subalign}[1][]{%
\begin{subequations}\begin{align}
  \BODY
\end{align}\label{#1}\end{subequations}
}

\NewEnviron{spliteq}{%
\begin{equation}\begin{split}
  \BODY
\end{split}\end{equation}
}

\newcommand{\cm}{\text{cm}}

\newcommand{\vecx}{\mathbf{x}}

\newcommand{\vecR}{\mathbf{R}}

\newcommand{\cdd}{{\cdot\cdot}}

\newcommand{\dvrsum}[2]{\sum\limits_{#1={-}#2/2}^{#2/2-1}}

\newcommand{\idx}{\text{idx}}

\newcommand{\RR}{\mathbb{R}}
\newcommand{\CC}{\mathbb{C}}
\newcommand{\HH}{\mathcal{H}}

\newcommand{\Dil}[2]{D_{#1,#2}}

\newcommand{\maxplus}{\mathbin{\stackrel{\text{max}}{+}}}

\newcommand{\inner}[2]{\langle#1,#2\rangle}
\newcommand{\maxinner}[2]{\inner{#1}{#2}_{\text{max}}}

\begin{document}

\title{Efficient few-body calculations in finite volume}

\author{S K\"onig}

\address{Department of Physics, North Carolina State University,
Raleigh, NC 27695, USA}

\ead{skoenig@ncsu.edu}

\begin{abstract}
Simulating quantum systems in a finite volume is a powerful theoretical tool to
extract information about them.
Real-world properties of the system are encoded in how its discrete energy
levels change with the size of the volume.
This approach is relevant not only for nuclear physics, where lattice methods
for few- and many-nucleon states complement phenomenological shell-model
descriptions and \textit{ab initio} calculations of atomic nuclei based on
harmonic oscillator expansions, but also for other fields such as simulations of
cold atomic systems.
This contribution presents recent progress concerning finite-volume simulations
of few-body systems.
In particular, it discusses details regarding the efficient numerical
implementation of separable interactions and it presents eigenvector
continuation as a method for performing robust and efficient volume
extrapolations.
\end{abstract}

\section{Introduction}

Simulating quantum systems in finite volume (FV), such as a cubic box with
periodic boundary conditions, is a well established theoretical approach to
extract information about them, going back to the early work of
Lüscher~\cite{Luscher:1985dn,Luscher:1986pf,Luscher:1990ux} who showed that
the real-world (\ie, infinite-volume) properties a the system are encoded
in how its (discrete) energy levels change as the size of the volume is varied.
The method has become a standard approach for example in Lattice Quantum
Chromodynamics (LQCD) to extract scattering information for hadronic systems,
and extending it in different directions, in particular to three-body systems,
is an area of active
research~\cite{Kreuzer:2010ti,Kreuzer:2012sr,Polejaeva:2012ut,Briceno:2012rv,%
Kreuzer:2013oya,Meissner:2014dea,Hansen:2015zga,Hammer:2017uqm,%
Hammer:2017kms,Mai:2017bge,Doring:2018xxx,Pang:2019dfe,Culver:2019vvu,%
Briceno:2019muc,Romero-Lopez:2019qrt,Hansen:2020zhy,Muller:2021uur}.
Moreover, few-body approaches formulated in FV can be used to match and
extrapolate LQCD results to an effective field theory (EFT)
description~\cite{Barnea:2013uqa,Detmold:2021oro,Kirscher:2015yda,%
Yaron:2022rmb}.

Bound-state energy levels have an exponential dependence on the size $L$ of the
periodic box that encodes asymptotic properties of the
state's wavefunction in infinite
volume~\cite{Luscher:1985dn,Konig:2011nz,Konig:2011ti}.
For a general bound state of $N\geq2$ particles with lowest breakup channel
into two clusters with $A$ and $N-A$ particles, respectively, the volume
dependence of the binding energy has been found to
be~\cite{Konig:2017krd,Konig:2020lzo}
\begin{equation}
 \Delta B_N(L) \equiv B_N(L) - B_N(\infty)
 = \frac{(-1)^{\ell+1} \sqrt{\tfrac{2}{\pi}}
 f(d)\abs{\gamma}^2}{\mu_{A|N-A}}
 \kappa^{2-d/2}_{A|N-A}L^{1-d/2} K_{d/2-1}(\kappa_{A|N-A}L) \,,
\label{eq:ANC-FV}
\end{equation}
where $d$ denotes the number spatial dimensions, $f(d)$ is a normalization
factor, $K_{d/2-1}$ is a modified Bessel function, and
\begin{equation}
 \kappa_{A|N-A} = \sqrt{2\mu_{A|N-A}(B_{N}-B_{A}-B_{N-A})}
\end{equation}
with the reduced mass $\mu_{A|N-A}$ of the two-cluster system.
Moreover, $\gamma$ is the asymptotic normalization coefficient of the cluster
wavefunction, a quantity that plays an important role for the description of
low-energy capture processes.
Equation~\eqref{eq:ANC-FV} implies that both $\gamma$ and $\kappa_{A|N-A}$ can
be extracted by fitting the volume dependence of numerical simulations.
In that regard it should be noted that most systems in nuclear physics of
practical interest feature more than one proton, and therefore
Eq.~\eqref{eq:ANC-FV} does not directly apply because it assumes pure
short-range interactions between all particles.
Work that derives the volume dependence for charged-particle bound states, \ie,
including the long-range Coulomb interaction, is nearly concluded at the time
this contribution is being written~\cite{Yu:2022xx}.
The following sections discuss recent progress regarding the efficient
numerical implementation of few-body systems in finite volume.
The material is based primarily on Refs.~\cite{Dietz:2021haj,Yapa:2022nnv}, but
includes additional details in particular in Sec.~\ref{sec:SepInt}.

\section{Discrete Variable Representation}

Few-body calculations in periodic finite boxes can be implemented elegantly
with a ``discrete variable representation (DVR)'' based on plane-wave states.
This method has been used and described in
Refs.~\cite{Klos:2018sen,Konig:2020lzo,Dietz:2021haj,Yapa:2022nnv}.
The following discussion elaborates on the use of separable interactions within
this framework, first introduced in Ref.~\cite{Dietz:2021haj}, with focus here
on an efficient numerical implementation.
To that end, we keep the general introduction of the method brief (referring
the reader to the papers cited above), but provide previously unpublished
details regarding the computation.

The DVR construction starts from plane-wave states defined on an interval
$L$,
\begin{equation}
 \phi_j^{(L)}(x) = \frac{1}{\sqrt{L}} \exp\left(\ii\frac{2\pi j}{L} x\right) \,,
\label{eq:PW-basis}
\end{equation}
with $j={-}N/2,\cdots N/2-1$ for even number of modes $N>2$ and where $x$
denotes the relative coordinate describing a two-body ($n=2$) system in one
dimension ($d=1$).
Given a set of equidistant points $x_k \in [{-}L/2,L/2)$ and weights $w_k = L/n$
(independent of $k$), DVR states are constructed as~\cite{Groenenboom:2001web}
\begin{equation}
 \psi_k(x) = \dvrsum{j}{N} \mathcal{U}^*_{kj} \phi_j(x) \,,
\label{eq:psi-dvr}
\end{equation}
with $\mathcal{U}_{ki} = \sqrt{w_k} \phi_i(x_k)$ defining a unitary matrix.
The index $k$ in Eq.~\eqref{eq:psi-dvr} covers the same range of integers as
the $j$ labeling the original plane-wave modes, and $\psi_k(x)$ is a
wavefunctions peaked at $x_k$.
Effectively, the plane-wave DVR can be thought of as a lattice discretization
that maintains the exact continuum energy-momentum dispersion relation.

Let now $\mathcal{B}_N^{(n,d)} = \{\ket{s}\}$ denote a basis of DVR states for
$A$ particles in $d$ spatial dimensions, with truncation parameter $N$.
In the following, the superscript $(n,d)$ is dropped to simplify the
notation.
A basis state $\ket{s}$ can then be written as
\begin{equation}
 \ket{s}
 = \ket{
  (k_{{1,1}},\cdd k_{{1,d}}),\cdd ,(k_{{n-1,1}},\cdd k_{{n-1,d}});
  (\sigma_1,\cdd \sigma_n)
 } \,,
\label{eq:s}
\end{equation}
where the $\sigma_i$ label optional spin degrees of freedom.
The formalism can be extended in a straightforward way to include additional
discrete degrees of freedom such as isospin.
In coordinate-space representation, where we use $\underline{x}$ to
collectively denote all relative
coordinates, the spatial part of a DVR state $\ket{s}$ is a tensor product of
one-dimensional DVR wavefunctions:
\begin{equation}
 \psi_s(\underline{x})
 = \braket{\underline{x}}{s}
 = \prod_{
  \substack{i=1,\cdot\cdot n-1 \\c=1,\cdot\cdot d}
 } \psi_{k_{i,c}}(x_{i,c}) \,.
\end{equation}

\subsection{Separable interactions}
\label{sec:SepInt}

Separable interactions, written, in their simplest form as rank-1 operators
$V = C \ket{g}\bra{g}$ have a number of desirable properties.
In particular, they are a popular choice for short-range interactions arising
in EFTs because they allow for closed-form
algebraic solutions for two-body bound-state and scattering calculations,
simplifying considerable the renormalization procedure that fixes, for given
choice of $\ket{g}$, the interaction strength---called ``low-energy constant
(LEC)'' in EFT context and denoted as $C$ above---to some physics input.
In coordinate space one obtains $V(x,x') = C \, g(x) g(x')$ with the ``form
factor'' $\braket{x}{g} = g(x)$.
In EFT, $g(x)$ is usually referred to as ``regulator'' because it incorporates
a momentum cutoff scale $\Lambda$, e.g., in the form of a Gaussian
function $\braket{x}{g} = g(x) = \exp\left({-}{\Lambda^2x^2}/{4}\right)$.
More generally, $\ket{g}$ can be conveniently chosen as a (super-)Gaussian
function in momentum space, $g(p) = \braket{x}{g} =
\exp\left({-}(p/\Lambda)^n\right)$ with $n = 2,4,6,\ldots$, and then $g(x)$ is
obtained from $g(p)$ by a Fourier transformation.

For simplicity the following discussion is restricted to one spatial dimension
since everything carries over to $d>1$ dimensions in a straightforward way.
For a two-body state, $\underline{x} = x_1 \equiv x$ and $\ket{s} =
\ket{k}$ with $k$ the single integer index that describes the DVR state.
Applying $V$ to an arbitrary linear combination
\begin{equation}
 \psi_s(\underline{x})
 = \braket{\underline{x}}{s}
 = \prod_{
  \substack{i=1,\cdot\cdot n-1 \\c=1,\cdot\cdot d}
 } \psi_{k_{i,c}}(x_{i,c})\
\end{equation}
is straightforward:
\begin{spliteq}
 \mbraket{s}{V}{\psi}
 &= \int\!\dd x \int\!\dd x' \, \psi_{s}^*(x') V(x,x')
 \sum_{s'\in B_N} c_{s'} \psi_{s'}(x') \\
 &= C \, g(x_{s}) \sum_{s'\in \mathcal{B}_N} c_{s'} \, g(x_{s'}) \,.
\label{eq:s-V-psi}
\end{spliteq}
For the general case ($n>2$ interacting particles), there can be interactions
$V_{ij}$ between any pair $(i,j)$ of particles.
This $V_{ij}$ then acts as an operator on the entire space, but it must
include appropriate Kronecker deltas for the ``spectator'' particles other than
the given pair $(i,j)$.
Schematically, Eq.~\eqref{eq:s-V-psi} becomes:
\begin{equation}
 \mbraket{s}{V_{ij}}{\psi}
 = \mathcal{N} \times C \, g(x_{s;ij}) \;
  \sum_{
    \mathclap{
     \substack{s'\in B_N \\
      r_{s';k,ij} = r_{s;k,ij} \forall k \neq
      i,j
     }
    }
   } \;
   c_{s'} \, g(x_{s';ij})
   \rule{0pt}{2em}
   \,.
\label{eq:s-V-psi-ij}
\end{equation}
The factor $\mathcal{N} = 2^d({L}/{n})^{d/2}$ arises as normalization from the
integral over spectator coordinates and $x_{s;ij}$ denotes the relative
distance (modulo the periodic boundary) between particles $i$ and $j$ in
configuration $\ket{s}$.
Moreover, $r_{s;k,ij}$ is the coordinate of particle $k$ relative
to the center of mass of the pair $(i,j)$.
This coordinate is needed to express the Kronecker deltas for spectators,
written as restriction on the sum over $s'$ in Eq.~\eqref{eq:s-V-psi-ij}.
As detailed in the following, with careful thought this relatively
complicated summation can be carried out efficiently in a numerical DVR
calculation.

\begin{figure}[htbp]
 \centering
 \includegraphics[width=0.4\textwidth]{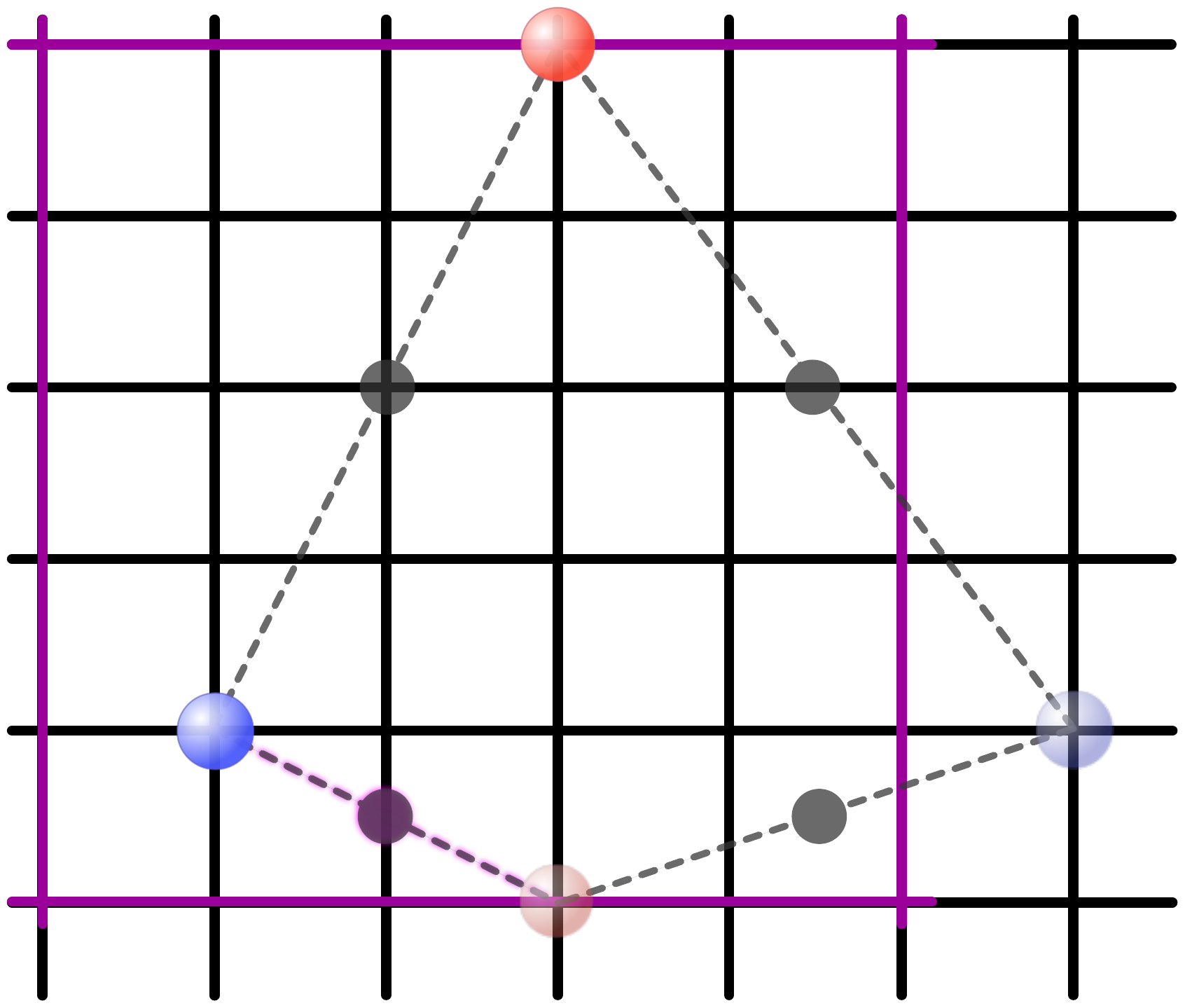}
 \caption{%
 Illustration of the center-of-mass ambiguity for two particles on
 a 2D lattice.
 Due to the periodic boundary condition, there are multiple points that can
 be considered the center of mass of the system that arise from taking into
 account shifted copies of the individual particles.
 As discussed in the text, the ambiguity can be lifted by picking as the center
 of mass that point among the different possibilities which minimizes the
 sum of distances between the particles and the center of mass (the lower left
 candidate in the figure).
 This construction generalizes in a straightforward way to any number of
 dimension and particles.
 \label{fig:periodic_com}
 }
\end{figure}

Calculating the $r_{s;k,ij}$ amounts to a partial transformation to a
particular set of Jacobi coordinates.
While this is straightforward in principle, special care has to be taken to
consistently define the center of mass of a cluster of particles in a box with
periodic boundary conditions.
As illustrated for a 2D system of two particles in Fig.~\ref{fig:periodic_com},
shifted copies of the individual particles introduce more than one point that
can be considered the center of mass of the system.
One way to uniquely and consistently define the center of mass of an $A$-body
cluster is to pick the point among all ``candidates'' that minimizes the sum
of distances of all particles measured with respect to the center of mass:
\begin{equation}
 \vecR_\cm = \argmin_{\vecR\in S_\cm(C)}\left(
  \sum_{j=1}^A d_L(\vecR,\vecx_j)^2
 \right) \,,
 \label{eq:R-cm}
\end{equation}
where $S_\cm(C)$ is the set of all possible center-of-mass coordinates for the
given configuration $C$ and $d_L$ measures the distance between two points as
the shortest path between them while accounting for the periodic boundary
condition.
This definition with $A=2$ is used to implement the spectator constraints in
Eq.~\eqref{eq:s-V-psi-ij}.

To construct now the numerical implementation of Eq.~\eqref{eq:s-V-psi-ij},
first note that the DVR basis states $\ket{s}$ as defined in Eq.~\eqref{eq:s}
can be mapped onto a set of integers with a straightforward prescription:
\begin{equation}
 \idx(\ket{s}) = \sum_{
  \substack{i=1,\cdot\cdot n-1 \\c=1,\cdot\cdot d}
 } (k_{i,c} + N/2) \times N^{i\times d+c}
 + \dim_{\text{spat}}(B_N) \times \idx_{\text{spin}}(\ket{s})
\label{eq:idx-s}
\end{equation}
Here $\dim_{\text{spat}}(B_N)$ denotes the pure spatial dimension of the DVR
basis (which is the maximum spatial index plus one, since counting according to
Eq.~\eqref{eq:idx-s} starts at zero), and $\idx_{\text{spin}}(\ket{s})$ denotes
an index, defined analogously to the first term in Eq.~\eqref{eq:idx-s}, for
the spin configuration $(\sigma_1,\cdot\cdot\sigma_n)$ alone.
This mapping is straightforward to invert in order to recover the entries
describing the state from $\idx(\ket{s})$.
All that is needed to extract a particular $k_{i,c}$ is a division with the
corresponding power of $N$.
Likewise it is possible to extract the spin and spatial parts as the integer
quotient or remainder of $\idx(\ket{s})$ with respect to
$\dim_{\text{spat}}(B_N)$.
It should however be noted that integer division is a relatively expensive
operation on a CPU.
Since the DVR calculation repeatedly requires many such operations in order to
work with states represented by integers, it is desirable to speed up this
aspect or the implementation with a fast integer-division ``trick.''
This procedure makes use of finite-width integer arithmetics to replace
division with a known divisor by a multiplication with a constant and a bit
shift, leaving only a one-time cost (for each desired divisor) to set up the
constant and shift.
As particular choice of algorithm, the code described here makes use of
\texttt{int\_fastdiv}~\cite{int-fastdiv}.

A mapping completely analogous to what is described in Eq.~\eqref{eq:idx-s} for
the basis states can be applied to translate the $r_{s;k,ij}$ to integer
numbers based on their components.
The only additional aspect that should be noted is that the center of mass for
a cluster of $A$ particles living on a lattice with spacing $a$ (given by $a =
L/N$) will
fall on a finer lattice with reduced spacing $\tilde{a} =
a/A$~\cite{Elhatisari:2017eno}.
The components of these coordinates can be expressed as integers ranging from
${-}AN/2$ to $AN/2$.
For the pairs appearing in the separable two-body interactions, $A=2$,
so in order to calculate indices for the $r_{s;k,ij}$ one needs to use a version
of Eq.~\eqref{eq:idx-s} with $N \to 2N$.
The spectator Kronecker deltas are then easily implemented by associating, for
each pair $(i,j)$,\footnote{%
For identical particles it suffices to consider explicitly just a single
representative pair because the interactions between all others will be
generated by the (anti-)symmetrization procedure.}
an integer with each state $\ket{s}$.
The sum in Eq.~\eqref{eq:s-V-psi-ij} is then carried out by summing for a given
$\ket{s}$ over those $\ket{s'}$ that have the same associated integer.
Numerically, this can be done efficiently as follows:

\paragraph{Preparation (one time cost)}
\begin{enumerate}
\item For each relevant pair, the center-of-mass index is calculated for each
 basis state and stored in an integer vector (array).
\item A copy of this vector is made and subsequently sorted, modifying the
 copy and storing in addition the permutation (another integer vector) that
 sorts the original vector.
\item The vectors together are used to construct a data structure representing
 the ``separable overlaps'' for each center-of-mass index.
 To that end, a loop is run over the sorted vector that counts how many
 subsequent entries from any given position have the same value.
 In the newly constructed date structure it associates with each range of equal
 values the corresponding range \emph{in the permutation vector} (which labels
 states in the basis).
 This latter range is again sorted in ascending order.\footnote{%
 This additional sorting is not strictly necessary, but it improves memory
 access patters for the subsequent potential application.}
 The result is a table describing sets of basis states with common
 center-of-mass index, keyed by that index.
\end{enumerate}
The values of the separable form factors $g(x_{s,ij})$ should ideally also be
precalculated for all states $\ket{s}$.
Having those values available can be used for an additional optimization that
sets the center-of-mass indices to a sentinel value in the first step above so
that irrelevant terms can be skipped entirely.

\paragraph{Application (repeated cost)}

A calculation of finite-volume energy levels, \ie, eigenstates of the
Hamiltonian expressed in the DVR basis, is performed iteratively with the
Lanczos/Arnoldi algorithm, using PARPACK~\cite{PARPACK-ng}.
This involves, for the potential part of the Hamiltonian, repeated execution of
Eq.~\eqref{eq:s-V-psi-ij}.
With the preparation work described above, this can now be done efficiently as
follows:
\begin{enumerate}
 \item For state $\ket{s}$, represented by its index calculated according to
 Eq.~\eqref{eq:idx-s}, the center-of-mass index is looked up in the first vector
 described in preparation step (i).
 \item The associated overlap data within the structure described in
 preparation step (iii) is located with a binary search, which is a fast
 operation.
 \item The sum in Eq.~\eqref{eq:s-V-psi-ij} is calculated as a loop, starting
 a the entry located in step (ii), proceeding as long as there are subsequent
 entries with the same center-of-mass index.
 This procedure becomes particularly efficient if the corresponding form-factor
 values are stored directly within the overlap data structure.
\end{enumerate}

In summary, the key design idea for an efficient application of separable
two-body potentials within the DVR basis is the use of sorted vectors to
represent the necessary data in a particular compressed format, exploiting that
binary (lower bound) searches can be used to look up entries with only
logarithmic complexity.
Reducing the memory footprint of the implementation this way is a tradeoff that
pays of particularly well for large-scale calculations (where overall memory is
a constraint), as well as for GPU-based implementations that benefit greatly
from limiting the amount of data that needs to be transferred between GPU and
CPU.

In practice there are other aspects concerning the distribution of the
implementation across multiple nodes in a computing cluster that complicate the
algorithm, as well as opportunities for further optimization of the procedure
(such as using compressed integer vectors for some of the data structures).
While those are beyond the scope of the present paper, hopefully the details
described above are useful to the interested reader.
In that regard, it should be pointed out that the design of the algorithm
is not specific to the DVR method, but applies with little or no modification
also to other methods based on lattice discretization.

\section{Fast volume extrapolation}

As mentioned in the introduction, finite-volume calculations carry information
about real-world properties of physical systems in how energy levels
\emph{change} as the size of the volume is varied.
This means that in general one does not want to merely run a calculation at
some fixed large volume, but across a range of volumes.
That is particularly true for extractions of asymptotic normalization
coefficients from the volume dependence of cluster
states~\cite{Konig:2011nz,Konig:2017krd} and for studying few-body resonance
states using finite-volume energy levels~\cite{Klos:2018sen}.

Running a single finite-volume calculation can come with a substantial
numerical cost, especially when one is interested in studying few-body systems
in large boxes~\cite{Dietz:2021haj}.
Recent work~\cite{Yapa:2022nnv} introduced a novel method to significantly
speed up calculations across a range of volumes by extending the technique of
``eigenvector continuation (EC)'' to extrapolate simulations from a given set
of volumes with exact results to a set of target volumes.
Each such ``emulated'' simulation has a cost that is greatly reduced compared
to an exact calculation in the same volume.

Eigenvector continuation was first introduced in Ref.~\cite{Frame:2017fah}
as a method to robustly extrapolate parametric dependencies of a
Hamiltonian $H = H(c)$ to a given target point $c_*$ from ``training data''
that may be far away from that point.
This technique is powerful yet simple and is based on exploiting information
contained in eigenvectors in the training regime.
Recent work~\cite{Bonilla:2022rph,Melendez:2022kid} has established EC as a
particular reduced-basis (RB) method that falls within a larger class of
model-order reduction (MOR) techniques.

Reference~\cite{Yapa:2022nnv} introduced an extension of EC called
``finite-volume eigenvector continuation (FVEC)'' that can be used to
extrapolate properties of quantum states calculated in a set of periodic boxes
with sizes $L_i$, $i=1,\ldots,N$ to a target volume $L_*$.
What is new in this case compared to standard EC is the parametric dependence
now appears directly within the basis, rather than in the Hamiltonian expressed
in a fixed basis for different values of the parameter.
Specifically, FVEC uses as training data states $\ket{\psi_{L_i}}$
at volumes $L_i$.\footnote{%
In a practical calculation one would generally consider sets of states
$\{\ket{\psi_{L_i}^{(j)}},\,j=1,\ldots,N_i\}$ at each volume $L_i$.
Conceptually this is no significant complication, so we avoid it here in order
to keep the notation simple.}
Naively, performing an EC extrapolation to a target volume $L_*$ then amounts
to constructing Hamiltonian and norm matrices, $H(L_*) = \big(H_{ij}(L_*)\big)$
and $N = \big(N_{ij}\big)$, respectively, with
\begin{subalign}[eq:H-N-naive]
 H_{ij}(L_*) &= \mbraket{\psi_{L_i}}{H_{L_*}}{\psi_{L_j}} \,, \\
 N_{ij}      &= \braket{\psi_{L_i}}{\psi_{L_j}} \,,
\end{subalign}
and solving generalized eigenvalue problem $H(L_*)\ket{\psi} = \lambda
N\ket{\psi}$.

The $H_{L_*}$ above denotes the Hamiltonian represented in a finite volume of
size $L_*$, and clearly it needs to be explained how Eqs.~\eqref{eq:H-N-naive}
can make sense at all given that the various states and operators are all
defined in different spaces.
The resolution developed in Ref.~\cite{Yapa:2022nnv} is to define a combined
space of periodic functions with different period by means of a union
\begin{equation}
 \HH = \bigcup_{\{L>0\}} \HH_L \,,
\end{equation}
where $\HH_L$ is the Hilbert space of functions $f: \RR \to \CC$ with period
$L$ (the 1D construction presented here for simplicity trivially generalizes to
$d>1$ dimensions).
With the standard (pointwise) addition and inner product, $\HH$ is not a vector
space: the pointwise sum of two periodic functions is in general not periodic,
and inner products defined as integrals have different integration domains for
two functions with different periods.
However, Ref.~\cite{Yapa:2022nnv} shows in detail how dilatation operations,
\begin{equation}
 (\Dil{L}{L'}f)(x) = \sqrt{\frac{L}{L'}}\,f\!\left(\frac{L}{L'}x\right) \,,
\end{equation}
can be used to match the period of any two (or more) functions, and then
the operations
\begin{equation}
 (f \maxplus g)(x) = (\Dil{L}{L'}f)(x) + g(x)
\end{equation}
and
\begin{equation}
 \maxinner{f}{g} = \inner{\Dil{L}{L'}f}{g}_{\HH_{L'}}
 = \int_{{-}L'/2}^{L'/2} {(\Dil{L}{L'}f)(x)}^* g(x) \, \dd x
\label{eq:maxinner}
\end{equation}
turn $\HH$ into a proper vector space with inner product.
Defined on this space, the matrix elements in Eqs.~\eqref{eq:H-N-naive} have
a well defined meaning.
%
%%%%%%%%%%%%%%%%%%%%%%%%%%%%%%%%%%%%%%%%%%%%%%%%%%%%%%%%%%%%%%%%%%%%%%%%%%%%%%
\begin{figure}[t!] \centering
\includegraphics[width=0.6\textwidth]{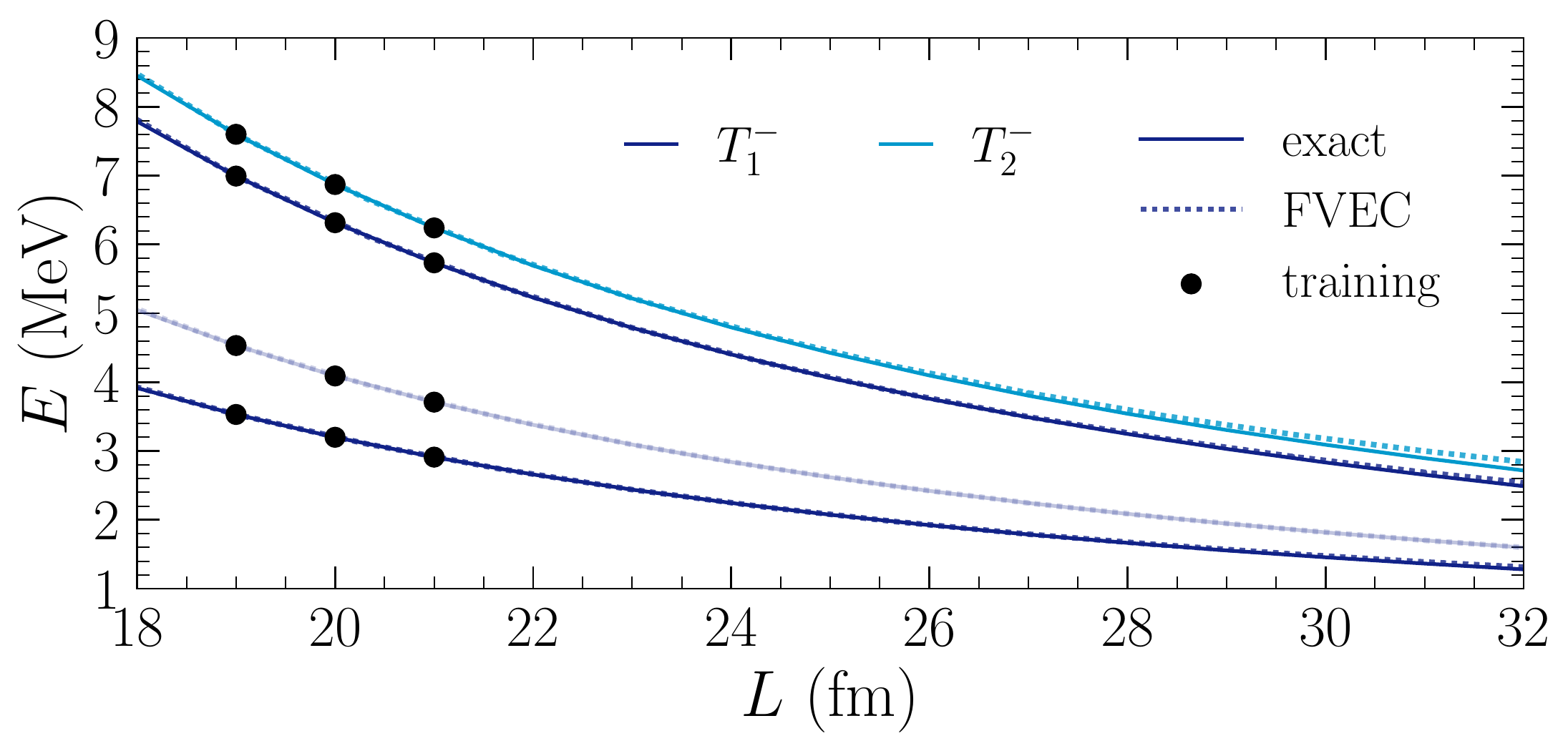}
 \caption{%
  Negative-parity $S_z=1/2$ finite-volume energy spectrum of three neutrons
  interacting via a separable contact potential fit to reproduce the
  neutron-neutron scattering length $a_{nn} = {-}18.9~\fm$.
  Solid lines show the exact states
  calculated in DVR bases with $N \leq 22$, whereas dashed lines indicate FVEC
  results obtained based on $N=22$ training data at three different box sizes
  (solid circles).
  A total number of $3\times8=24$ training data were used to generate this plot,
  covering a subset of states from four three-dimensional cubic group
  multiplets.
  \label{fig:En-3n-SepContact-n4-250-phys-Sz12-Pm-EC}
 }
\end{figure}
%%%%%%%%%%%%%%%%%%%%%%%%%%%%%%%%%%%%%%%%%%%%%%%%%%%%%%%%%%%%%%%%%%%%%%%%%%%%%%
%
Moreover, for truncated bases of plane-wave states as defined in
Eq.~\eqref{eq:PW-basis}, $S_{L,N} = \{ \phi_j^{(L)} : j = 1,\cdd N \}$,
 $\Dil{L}{L'}$ defines a bijection between $S_{L,N}$ and $S_{L',N}$.
Therefore, if $\psi$ and $\psi'$ are functions expanded upon $S_{L,N}$
and $S_{L',N}$, respectively, taking the inner product of their coefficient
vectors in $\RR^N$ is equivalent to considering the inner product on $\HH$ as
defined in Eq.~\eqref{eq:maxinner}.
This observation, which carries over directly to the DVR basis (because that is
unitary rotation of the plane-wave basis), provides a straightforward practical
implementation.

An application of the method is shown in
Fig.~\ref{fig:En-3n-SepContact-n4-250-phys-Sz12-Pm-EC} for a system of three
neutrons interacting via an $n=4$ super-Gaussian separable potential in the
(finite-volume analog of the) ${}^1S_0$ channel, with low-energy constant $C$ is
fixed to reproduce the $nn$ scattering length $a_{nn} = {-}18.9~\fm$.
Using training data from three $N=22$ DVR calculations at $L=19,20,21~\fm$,
FVEC accurately extrapolates energy levels of low-lying negative-parity states
(falling into two different representations, $T_1^-$ and $T_2^-$ of the cubic
symmetry group~\cite{Johnson:1982yq}) over a large range of volumes.
As emphasized numerical cost of the FVEC extrapolation is greatly reduced
compared to exact calculations over the entire range of volumes shown in
Fig.~\ref{fig:En-3n-SepContact-n4-250-phys-Sz12-Pm-EC}.
Finite-volume eigenvector continuation is therefore an important technological
advancenment that will enable the study of larger few-body systems in FV, such
as for example searches for a tetraneutron resonance state.

\ack
Parts of this work have been presented at the 13th International Spring Seminar
on Nuclear Physics, held in Sant’Angelo d’Ischia in May 2022.
I would like to thank the organizers of that meeting for hosting a stimulating
conference in a beautiful environment, and for giving me the opportunity to
contribute.
Moreover, I would like to thank my co-authors on the papers that this
contribution is based on in part.
This work was supported in part by the National Science Foundation under Grant
No. PHY--2044632.
This material is based upon work supported by the U.S. Department of Energy,
Office of Science, Office of Nuclear Physics, under the FRIB Theory Alliance,
Award No. DE-SC0013617.
Computing resources for the results presented in this work were provided by the
Jülich Supercomputing Center and by the Henry2 high-performance computing
cluster operated by North Carolina State University.

\section*{References}

\bibliographystyle{iopart-num}
\providecommand{\newblock}{}

\end{document}